\documentclass{article}

\usepackage{PRIMEarxiv}
\usepackage[utf8]{inputenc} 
\usepackage[T1]{fontenc}    
\usepackage{hyperref}       
\usepackage{url}            
\usepackage{booktabs}       
\usepackage{amsfonts}       
\usepackage{nicefrac}       
\usepackage{microtype}      
\usepackage{fancyhdr}       
\usepackage{graphicx}       
\graphicspath{{media/}}     

\title{PersonaAI: Leveraging Retrieval-Augmented Generation and Personalized Context for AI-Driven Digital Avatars}

\author{
  Elvis Kimara, Kunle S. Oguntoye, Jian Sun \\
  Iowa State University \\
  Ames, IA, USA \\
  \texttt{\{ekimara, oguntoye, jsun29\}@iastate.edu} \\
}

\begin{document}
\maketitle

\begin{abstract}
This paper introduces PersonaAI, a cutting-edge application that leverages Retrieval-Augmented Generation (RAG) and the LLAMA model to create highly personalized digital avatars capable of accurately mimicking individual personalities. Designed as a cloud-based mobile application, PersonaAI captures user data seamlessly, storing it in a secure database for retrieval and analysis. The result is a system that provides context-aware, accurate responses to user queries, enhancing the potential of AI-driven personalization.

Why should you care? PersonaAI combines the scalability of RAG with the efficiency of prompt-engineered LLAMA3, offering a lightweight, sustainable alternative to traditional large language model (LLM) training methods. The system’s novel approach to data collection, utilizing real-time user interactions via a mobile app, ensures enhanced context relevance while maintaining user privacy. By open-sourcing our implementation, we aim to foster adaptability and community-driven development.

PersonaAI demonstrates how AI can transform interactions by merging efficiency, scalability, and personalization, making it a significant step forward in the future of digital avatars and personalized AI.
\end{abstract}


\section{Introduction}  
Artificial intelligence (AI) has become a transformative force across numerous domains, fundamentally reshaping the way humans interact with technology. While AI-powered virtual assistants such as Siri, Alexa, and ChatGPT have excelled in handling routine tasks and providing general-purpose assistance, these systems often lack the capacity to personalize interactions to reflect individual traits, preferences, and nuances. This inherent limitation has created a significant gap in user-centric AI solutions, where the interaction feels impersonal, static, and contextually inadequate.

\textbf{PersonaAI} addresses this critical need by introducing a novel system capable of creating AI personas that mirror individual personalities. Designed to serve diverse user needs, PersonaAI offers applications ranging from preserving the digital legacy of loved ones to assisting professionals who require consistent and accurate information delivery in their absence. Unlike traditional large language models (LLMs), which rely heavily on users repeatedly providing context, PersonaAI employs \textbf{Retrieval-Augmented Generation (RAG)} to seamlessly capture, store, and retrieve user-specific data, enabling highly relevant, \textbf{context-aware, and personalized responses}.

The need for personalization is especially urgent in today’s fast-paced and interconnected world. Generic AI interactions fail to engage users on a deeper level, diminishing user satisfaction and productivity. PersonaAI bridges this gap by utilizing a \textbf{cloud-based mobile application} that captures user data in real-time, securely storing it in a cloud database. By leveraging scalable and efficient LLMs such as LLAMA3 and combining them with advanced \textbf{prompt engineering techniques}, PersonaAI ensures a lightweight and cost-effective solution without the computational overhead of traditional LLM training.

Beyond technical advancements, PersonaAI is built with a strong focus on \textbf{user trust and reliability}. Recognizing the challenges of AI hallucinations—generating responses that are incorrect or misleading—PersonaAI incorporates mechanisms to transparently acknowledge limitations. By explicitly responding with "I don't know" when appropriate, the system mitigates inaccuracies and enhances user confidence. This approach is aligned with the ethical considerations necessary for deploying AI in sensitive and user-centric applications.

This paper delves into the methodology, design, and implementation of PersonaAI, offering the following key contributions:  
\begin{enumerate}
    \item \textbf{A Novel Approach to Personalization}: PersonaAI combines RAG with LLAMA3 and advanced prompting to deliver tailored interactions that adapt dynamically to individual user profiles.  
    \item \textbf{Scalable and Efficient Architecture}: The system demonstrates the feasibility of using scalable LLMs with retrieval-based augmentation, significantly reducing the computational costs associated with training domain-specific AI models.  
    \item \textbf{Innovative Data Collection through a Mobile App}: A seamless and secure data ingestion mechanism captures user contexts while prioritizing privacy and usability.  
    \item \textbf{Ethical Design and Reliability}: PersonaAI enhances user trust through transparent handling of uncertainties, setting a benchmark for ethical AI design.  
\end{enumerate}

By exploring these contributions, this paper aims to advance the understanding and practical deployment of personalized AI systems, highlighting their potential to transform digital interactions. The results and insights from PersonaAI lay the groundwork for broader applications in personalized education, healthcare, and digital legacy preservation.

\section{Related Works}  
Personalized AI systems have garnered significant attention in recent years, driven by advancements in natural language processing (NLP), retrieval-augmented generation, and memory-based neural architectures. Despite these advancements, existing solutions often fall short in delivering deeply personalized and scalable interactions, presenting opportunities for innovation.

Large language models (LLMs) such as OpenAI's GPT series [1] have redefined the capabilities of AI systems in generating coherent and contextually appropriate text. GPT-3 and GPT-4, for example, exhibit exceptional performance in natural language understanding and generation. However, these models typically operate as general-purpose assistants, requiring users to repeatedly provide context to produce personalized responses. This dependence on manual input limits their effectiveness in scenarios demanding dynamic and user-centric personalization.

Retrieval-Augmented Generation (RAG), introduced by Facebook AI [2], addresses some of these limitations by combining external retrieval mechanisms with generative models. RAG enables the dynamic integration of context-specific information into the response generation process, significantly enhancing relevance and accuracy. Despite its promise, existing RAG implementations often focus on narrow, domain-specific applications, leaving a gap in its application to user-driven, personalized interactions at scale.

Efforts to develop personalized virtual assistants, such as Google’s Meena [3] and Amazon Alexa [4], have incorporated user preferences and intent recognition to tailor interactions. These systems rely on static user profiles and predefined intent libraries, which can hinder their ability to adapt to nuanced or evolving user behaviors. Furthermore, their personalization strategies are typically limited to specific use cases, lacking the flexibility to generalize across diverse user contexts.

Memory-augmented neural networks [5], such as those pioneered by Weston et al., have advanced the capability of AI systems to retain and recall historical user data. These networks facilitate long-term personalization by enabling systems to adapt to user preferences over time. However, challenges remain in scaling memory networks efficiently, particularly as the volume of stored user data grows, which can lead to computational bottlenecks and diminished response times.

Ethical considerations have emerged as a critical area of research in personalized AI. Studies by Binns et al. [6] and Jobin et al. [7] emphasize the importance of transparency, data privacy, and user consent in AI systems that handle sensitive personal information. The deployment of personalized AI systems must prioritize secure data handling and provide users with control over how their data is captured and utilized. These principles are crucial for fostering trust and ensuring the responsible use of AI.

Building on these foundational works, PersonaAI addresses key gaps in the field. By leveraging RAG in conjunction with LLAMA models, PersonaAI dynamically retrieves and utilizes user-specific contexts to generate accurate, personalized responses. Its innovative data collection approach, implemented through a cloud-based mobile application, ensures seamless and privacy-preserving context capture. Furthermore, PersonaAI explicitly addresses ethical concerns by embedding mechanisms to handle uncertainty and transparently acknowledge limitations, enhancing user trust and system reliability.

In summary, PersonaAI builds on the progress of previous works while introducing novel methodologies to achieve scalable, secure, and deeply personalized interactions. Its contributions demonstrate the potential of AI systems to transform applications in education, healthcare, and beyond, paving the way for broader adoption of personalized AI technologies.

\section{Experimental Platform}
This section details the infrastructure, algorithms, and deployment strategies employed in the project.

\subsection{Infrastructure and Equipment}  
The infrastructure for PersonaAI comprises both computational and application layers to ensure seamless operation. The key components include:  
\begin{itemize}  
    \item \textbf{Cloud Storage and Databases}: A secure cloud-based database is utilized for storing user-specific data, including embedded vectors, transcriptions, and metadata. Firebase was selected for its robust hosting and real-time database capabilities.  
    \item \textbf{GPU Instances}: Transformer models were deployed on NVIDIA GPU instances to support efficient training and inference, particularly for LLAMA 2 models.  
    \item \textbf{Mobile Application}: A react native mobile application was developed to capture user data through voice-to-text conversion. This app periodically records user interactions and uploads transcriptions to the cloud database, ensuring real-time data ingestion and processing.  
    \item \textbf{Backend Server}: A Python-based backend using Flask was implemented to handle user authentication, query processing, and context retrieval workflows. Docker containerization ensured modularity and scalability.  
\end{itemize}  

\subsection{Algorithms and Data Structures}  
User data, whether textual or speech-derived, is organized using a dictionary-based data structure with keys such as \texttt{text} (transcribed data), \texttt{timestamp}, \texttt{username}, and an \texttt{embedded\_vector}. The embedding vectors were generated using the Hugging Face BAAI/bge-small-en model, producing 384-dimensional representations optimized for similarity computation.

To process user queries, a cosine similarity function ranks stored data by relevance. The top $k$ (2–5) most relevant contexts are dynamically retrieved and appended to the prompt sent to the LLAMA model for response generation. This dynamic retrieval ensures contextual relevance while minimizing computational overhead.

\subsection{Experimentation}  
To evaluate the performance of PersonaAI, we adopted a Retrieval-Augmented Generation (RAG) architecture for synthesizing responses from the top-$k$ retrieved contexts. Experiments were conducted to compare the system's performance against several baselines:  
\begin{itemize}  
    \item \textbf{Baseline ChatGPT (No User Data)}: To establish a performance baseline for general-purpose AI systems.  
    \item \textbf{ChatGPT with User Data}: To assess the relevance and personalization capabilities of PersonaAI.  
    \item \textbf{LLAMA 2 Models}: Both the 13B and 70B parameter variants were evaluated to identify the optimal balance between computational efficiency and response quality.  
\end{itemize}  

The system was also enhanced to remember user contexts from previous interactions, improving the continuity and accuracy of responses. Performance metrics such as response latency, contextual accuracy, and user satisfaction were measured across these experiments.  

\subsection{Deployment Strategy}  
PersonaAI’s deployment was structured to support modularity, scalability, and ease of integration. The deployment architecture included:  
\begin{itemize}  
    \item \textbf{Backend Services}: The backend, developed in Python using Flask, manages API communication and facilitates the interaction between the database, retrieval mechanisms, and the LLAMA model. Docker was employed for containerization, ensuring portability and scalability across deployment environments.  
    \item \textbf{Frontend Interface}: In addition to the app, a frontend application, built with React TypeScript, serves as the user-facing interface, handling tasks such as user authentication, data input, and messaging. The frontend also manages integration with the backend for seamless query processing and data retrieval.  
    \item \textbf{Cloud Integration}: Firebase was utilized for its real-time synchronization and secure database management, ensuring low-latency interactions and reliable storage for user data.  
\end{itemize}  

This modular deployment approach ensures that PersonaAI can be scaled effectively while maintaining high performance and responsiveness across a diverse range of use cases.

\section{Methodology}

The methodology adopted in this paper is inspired by the Retrieval-Augmented Generation (RAG) workflow, optimized for personalization through efficient data structuring, contextual retrieval, and prompt engineering. The key components include data preparation, query processing, language model integration, and dataset evaluation, as detailed below.

\subsection{Data Preparation and Chunking Strategy}

Users’ data are indexed and embedded into 384-dimensional vectors using the pre-trained Hugging Face BAAI/bge-small-en model. Although larger models (e.g., with vector sizes up to 1024) offer enhanced performance, we opted for a computationally efficient approach by implementing a recursive character chunking strategy. This method divides user data into $n$-chunks with a maximum size of 200 characters and a 25\% overlap to ensure continuity between fragments. The chunking follows a hierarchical structure, splitting the data recursively by paragraph, line, space, and character as needed to meet the size constraints. Figure~\ref{fig:chunking} illustrates the chunking process.

\begin{figure}[h]
\centering
\includegraphics[width=\linewidth]{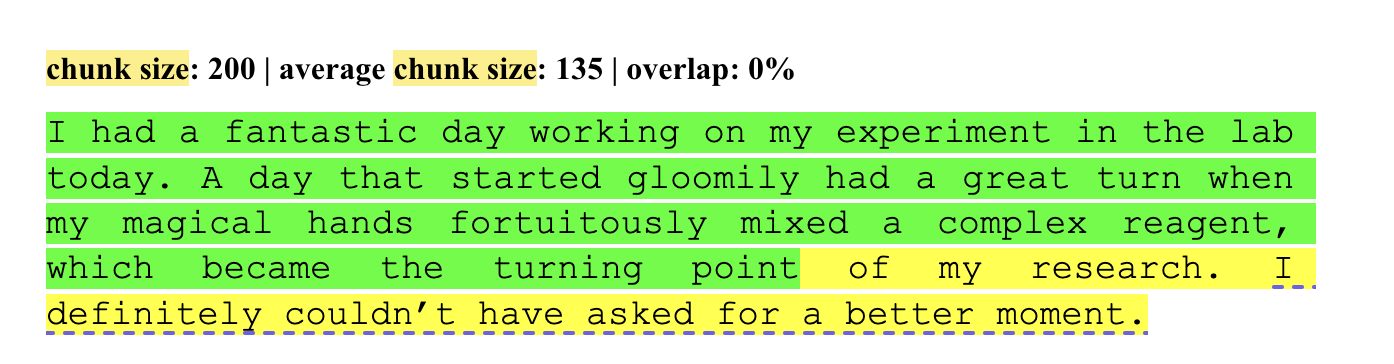}
\caption{Recursive character chunking strategy with 200-character size and 0\% overlap (demo).}
\label{fig:chunking}
\end{figure}

Each chunk is enriched with metadata, including \texttt{date}, \texttt{userID}, and the embedding vector generated by the BAAI/bge-small-en model. The 25\% overlap ensures continuity between chunks and minimizes contextual fragmentation, improving retrieval accuracy.

\subsection{Query Processing and Retrieval}

A robust context retrieval framework is critical to the success of any RAG system. To identify the most relevant data chunks for a user query, we implemented a dot product ranking mechanism based on cosine similarity. The retrieval workflow is as follows:
\begin{itemize}
    \item The user query is embedded using the same embedding model as the data.
    \item The embedded query vector is matched against the user’s database of size $N \times M$, where $N$ represents the total number of chunks and $M$ the vector size (384).
    \item The resulting similarity scores are ranked, and the top-$k$ most relevant chunks (where $k$ ranges from 2 to 5) are retrieved.
\end{itemize}

This retrieval process ensures computational efficiency for smaller databases. However, as the database grows, larger $k$ values may introduce irrelevant or redundant information, risking hallucinations or factually incorrect outputs. Optimizing $k$ remains a focus for future scalability improvements.

\subsection{Language Model Integration and Prompt Engineering}

The retrieved contextual data serves as input to a language model for generating personalized responses. For this study, we utilized LLAMA 2-70B, an open-source model with demonstrated capabilities in generating safe, reliable, and user-adaptive responses. LLAMA 2-70B was selected for its balance of scalability and performance, outperforming comparable models like ChatGPT-3.5 in generative tasks. We later proceeded to use LLAMA3. 

Effective prompt engineering was a cornerstone of our methodology. Prompts were tailored to different use cases to maximize the model’s contextual understanding and minimize errors:
\begin{itemize}
    \item \textbf{Use Case 1: New Users (Empty Database)}  
    For users without pre-existing data, prompts guide the model to provide generic, engaging, and honest responses or request additional context when necessary.
    \item \textbf{Use Case 2: Returning Users (Available Context)}  
    For users with stored data, prompts dynamically integrate retrieved contexts, enabling precise, personalized, and factually grounded responses.
\end{itemize}

The prompt templates ensure the model responds appropriately to general questions while acknowledging limitations by explicitly stating, “I DO NOT KNOW,” when information is unavailable.

\subsection{Datasets and Evaluation}

The framework was evaluated using two datasets: 
\begin{enumerate}
    \item \textbf{SpongeBob Dataset:} This dataset consists of dialogues from the SpongeBob SquarePants series, providing a proof-of-concept for creating AI personas based on linguistic quirks and informal speech patterns. While useful for testing, the dataset contains significant noise due to short responses and non-verbal expressions, limiting its utility without preprocessing.
    \item \textbf{Simulated College Journal:} To address the limitations of the SpongeBob dataset, a simulated dataset was curated, emulating real-world user interactions. This dataset includes emails, personal reflections, notes from therapy sessions, and summaries of daily activities, offering a more structured and realistic context for testing personalization capabilities.
\end{enumerate}

The datasets were used to evaluate the ability of LLAMA 2-70B to mirror user-specific quirks while maintaining the generative quality required of standard language models.

\subsection{Building a Secure SaaS Framework}

The success of the framework prompted the development of a SaaS platform where each user is assigned a unique account with robust authentication and data privacy safeguards. User data is securely stored and processed, ensuring compliance with ethical standards and privacy laws. This implementation demonstrates the potential for large-scale deployment of personalized AI systems in real-world scenarios.

\subsection{Key Innovations}

The methodology presented in this paper introduces several key innovations:
\begin{itemize}
    \item \textbf{Efficient Context Chunking:} A recursive strategy ensures data continuity and minimizes loss of contextual relevance.
    \item \textbf{Dynamic Retrieval:} Optimized retrieval algorithms strike a balance between computational efficiency and contextual accuracy.
    \item \textbf{Ethical Design:} Prompts are explicitly structured to mitigate hallucinations and foster user trust.
    \item \textbf{Real-World Adaptation:} A cloud-based SaaS implementation ensures scalability and secure personalization.
\end{itemize}

This comprehensive methodology establishes the foundation for scalable, ethical, and user-centric AI systems.

\section{Results}

The proposed framework was evaluated through various use cases, demonstrating its ability to generate accurate, context-aware, and personalized responses. By leveraging the computational efficiency and inference capabilities of three large language models (LLMs) — LLAMA-3.1-405B-Instruct, LLAMA-3-70B-Instruct, and LLAMA-2-70B-Chat — the proposed architecture performance was analyzed across different scenarios. This section presents the results, highlighting general knowledge performance, user-context-based testing, user feedback, and overall system evaluation.

\subsection{Use Cases and Model Performance}

\subsubsection{General Knowledge Use Case}

The first use case involved testing the models' ability to respond to general knowledge questions without user-specific contexts. These questions evaluated baseline knowledge capabilities and the handling of uncertainty or outdated information. A question and responses across the tested models are summarized below.

\label{lst:general_knowledge_use_case_1}

\begin{verbatim}
Question 1: Who is the president of Nigeria?

LLAMA-3.1-405B-Instruct:
As of my last update in 2021, the President of Nigeria is Muhammadu Buhari. However, please note that political positions can change, and I might not have the most current information. For the most accurate and up-to-date answer, I recommend checking the latest news or official government sources.

LLAMA-2-70B-Chat:
I do not know. I'm just an AI, I don't have access to real-time information and my knowledge cutoff is August 2022. The current president of Nigeria is Muhammadu Buhari, but I am not aware of any changes in leadership since my knowledge cutoff. However, I suggest checking with a reliable news source for the most up-to-date information.

LLAMA-3-70B-Instruct:
As of my knowledge cutoff, the President of Nigeria is Muhammadu Buhari. However, please note that this information may change over time, and I may not always have the most up-to-date information. If you need the most current information, I recommend checking a reliable news source or official government website for the latest updates.
\end{verbatim}

The results highlight the following:
\begin{itemize}
    \item \textbf{LLAMA-3.1-405B-Instruct}: 
    \item \textbf{LLAMA-2-70B-Chat}: 
    \item \textbf{LLAMA-2-70B-Instruct}: 
\end{itemize}

\subsubsection{General Recommendations (Non-Personalized) Use Case}
Non-personalized recommendations arise when the available contexts are not close to the query or the user's database is empty.

\begin{verbatim}
Question 1: Recommend a movie to watch this weekend?

LLAMA-3.1-405B-Instruct:
There are so many great movies out there, but I'll give you a classic recommendation that's hard to go wrong with.
Have you considered watching "The Shawshank Redemption" (1994) directed by Frank Darabont? It's a timeless drama film with outstanding performances, a gripping storyline, and a powerful message of hope and redemption. The movie has a high rating on IMDB and is widely regarded as one of the greatest films of all time.
If you're in the mood for something else, please give me a hint about your preferred genre (e.g., action, comedy, sci-fi, horror, etc.) or any specific actors you'd like to see, and I can try to suggest an alternative!

LLAMA-2-70B-Chat:
Sure, I'd be happy to help! However, I would need some more information to give a recommendation
that's tailored to your tastes. Here are a few questions to help me narrow down some options:
1. What type of movie are you in the mood for? (e.g. action, comedy, romance, horror, etc.)
2. Are you open to watching a classic film, or do you prefer something more recent?
3. Is there a particular actor or actress you'd like to see in the movie?
4. Do you have a preferred streaming platform or do you plan on renting/buying the movie?
Once I have a better understanding of your preferences, I can give you a personalized movie recommendation for the weekend!

LLAMA-3-70B-Instruct:
What a great question! There are so many amazing movies out there, but I'd love to know a bit more about your preferences to make a recommendation that's just right for you.
Do you have a favorite genre, like action, rom-com, or horror? Are you open to watching a classic film, or something more recent? Let me know, and I'd be happy to suggest a movie that fits your taste!
\end{verbatim}
The results highlight the following:
\begin{itemize}
    \item \textbf{LLAMA-3.1-405B-Instruct}: 
    \item \textbf{LLAMA-2-70B-Chat}: 
    \item \textbf{LLAMA-2-70B-Instruct}: 
\end{itemize}

\subsubsection{Personalized Recommendations Use Case}
Given the 5.1.2 is a non-personalized recommendation request and the models suggest user to provide their preferences, we added a simple text to the database;

New knowledge: "My colleagues at work would not stop talking about Anime on Netflix. I sure should watch one in the coming days."

\begin{verbatim}
Question 1: Recommend a movie to watch this weekend?

LLAMA-3.1-405B-Instruct:
It seems the context is about watching an Anime on Netflix, but the question asks for a movie recommendation. I'll provide a general response for movie recommendations, as it's unrelated to the context.
Here are some popular movie recommendations across various genres:
1. Action: "Mission: Impossible - Fallout" (2018)
2. Comedy: "The Hangover" (2009)
3. Drama: "Parasite" (2019)
4. Sci-Fi: "Inception" (2010)
5. Romance: "Crazy, Stupid, Love" (2011)
If you have a specific genre or theme in mind, I can give you more tailored suggestions!

LLAMA-2-70B-Chat:
 Sure, I can help you with that! Based on your colleagues' interest in Anime and your desire to watch a movie this weekend, I recommend checking out "Attack on Titan" on Netflix. It's a popular and highly-rated anime series that has gained a lot of attention in recent years. The show is set in a post-apocalyptic world where humans are under attack by giant humanoid creatures, and it explores themes of survival, humanity, and the struggle against monstrous oppressors.
If you're not interested in anime, there are plenty of other great movies and shows available on Netflix. Here are a few recommendations:
* "Stranger Things" - a sci-fi horror series that follows a group of kids as they battle supernatural forces in their small town.
* "The Umbrella Academy" - a superhero series that follows a dysfunctional family of adopted siblings with superpowers as they try to save the world from an impending apocalypse.
* "The Witcher" - a fantasy series based on the popular video game and book series, which follows a monster hunter named Geralt of Rivia as he navigates a magical world filled with political intrigue and mythical creatures.
I hope these recommendations help you find something you'll enjoy watching this weekend!

LLAMA-3-70B-Instruct:
I think I can help you with that!
Considering your colleagues' enthusiasm for Anime on Netflix, I'd like to recommend an Anime movie that's popular and well-received. Have you considered watching "Spirited Away" (2001) directed by Hayao Miyazaki? It's a timeless classic, widely acclaimed, and available on Netflix in many regions.
If you're not in the mood for Anime or would like an alternative, I can suggest a few other movies that are highly rated and suitable for a weekend watch. For instance, you might enjoy "The Shawshank Redemption" (1994), "The Grand Budapest Hotel" (2014), or "The Pursuit of Happyness" (2006). Let me know if you have any specific genre preferences, and I can provide more tailored recommendations!
\end{verbatim}
The results highlight the following:
\begin{itemize}
    \item \textbf{LLAMA-3.1-405B-Instruct}: 
    \item \textbf{LLAMA-2-70B-Chat}: 
    \item \textbf{LLAMA-2-70B-Instruct}: 
\end{itemize}

\subsubsection{Event Reminder Use Case}
This use case tests the model capability in understanding time-related events.

New knowledge: "Today is Monday, 23rd December 2024. I do have an appointment with my Doctor next week wednesday. Also, I would like to call some friends anytime next week."

\begin{verbatim}
Question 1: remind me what I am doing this week?
Question 2: create a to-do plan for me for next week

LLAMA-3.1-405B-Instruct:
Q1: You have an appointment with your Doctor next Wednesday. Additionally, you mentioned wanting to call some friends sometime next week, but there's no specific day or time scheduled for that.
Q2: Based on the context, I can help you create a to-do plan for next week. Here's a suggested plan:
**Next Week (December 30, 2024 - January 5, 2025)**
1. **Wednesday, January 1, 2025**:
        * Attend your scheduled doctor's appointment.
2. **Any day next week ( flexible)**:
        * Call your friends to catch up.
Additional suggestions to consider:
* Review your schedule for the upcoming week and make sure you're prepared for any other commitments or tasks.
* Make a list of any errands or chores you need to complete and allocate time for them.
* Consider setting aside some time for self-care and relaxation, especially during the holiday season.
Feel free to adjust this plan according to your specific needs and preferences!

LLAMA-2-70B-Chat:
Q1: Based on the information provided in the context, you have an appointment with your doctor next week on Wednesday, and you would like to call some friends sometime next week.
As for what you are doing this week, it appears that you have a doctor's appointment scheduled for next Wednesday. Additionally, you have expressed a desire to call some friends sometime during the upcoming week.
Is there anything else I can assist you with?
Q2: I'm not able to create a to-do plan for you for next week as I don't have access to your personal information and schedule. However, I can suggest some general tasks that you may want to consider adding to your to-do list for next week:
1. Call your friends: As you mentioned, you would like to call some friends next week. Consider scheduling specific times to call them and make sure to add their contact information to your list.
2. Prepare for your doctor's appointment: Make sure to prepare any necessary documents or information that you need to bring with you to your appointment. You may also want to schedule some time to review any questions or concerns you have for your doctor.
3. Plan your holiday activities: Next week is the holiday season, so you may want to plan some activities or events that you want to participate in. Consider adding tasks such as shopping for gifts, decorating your home, or scheduling time with family and friends.
4. Review your schedule: Take some time to review your schedule for the upcoming week and make sure that you have all the necessary information and materials for any appointments or events that you have planned.
5. Set reminders: Make sure to set reminders for any important tasks or deadlines that you have for next week. This will help ensure that you stay on track and don't forget anything important.
I hope this helps! Let me know if you have any other questions.

LLAMA-3-70B-Instruct:
Q1: This week, you don't have any specific appointments or plans mentioned in the context, except that you want to call some friends anytime next week. However, next week Wednesday, you have an appointment with your Doctor.
Q2: Here's a suggested to-do plan for you for next week:
Monday, 30th December 2024:
* Call friends to catch up (flexible timing)
Tuesday, 31st December 2024:
* Continue calling friends to catch up (flexible timing)
Wednesday, 1st January 2025:
* Attend doctor's appointment (scheduled)
* Call friends to catch up (flexible timing, if not already done)
Thursday, 2nd January 2025:
* Call friends to catch up (flexible timing, if not already done)
Friday, 3rd January 2025:
* Call friends to catch up (flexible timing, if not already done)
This plan allows you to attend your doctor's appointment on Wednesday and catch up with friends at your convenience throughout the week. Feel free to adjust the plan as needed to fit your personal schedule and preferences!

\end{verbatim}
The results highlight the following:
\begin{itemize}
    \item \textbf{LLAMA-3.1-405B-Instruct}: 
    \item \textbf{LLAMA-2-70B-Chat}: 
    \item \textbf{LLAMA-2-70B-Instruct}: 
\end{itemize}

\subsubsection{Writing Style Use Case}
This use case explores the ability of the models to mimic user's writing style. Here, a new knowledge is added, which typifies how a British writes emails to his friends. 

New knowledge: "Yo fam,
Was thinkin’ we hit up a flick later, yeah?
Bet it’s gonna be bare vibes, man – proper lit all day and night with this peng weather.
Safe, bruv. Catch ya soon.
Your mate, broski."

\begin{verbatim}
Question 1: using my past email writing style, write a letter to Paul, asking him for updates on our last conversation

LLAMA-3.1-405B-Instruct:
Blud,
Was thinkin' 'bout our last convo, yeah? Wanted to hit you up and see if there's any updates, fam? We was discussin' some pretty deep stuff, and I'm proper keen to hear how it's all goin' down, bruv.
Safe, Paul. Catch ya soon, and let me know what's good, aight?
Your mate, [Your Name

LLAMA-2-70B-Chat:
Yo Paul,
Hope you're doin' well, my dude! It's been a minute since we last spoke, and I was wonderin' if you had any updates on that thing we were discussin' last time?
I know you said you were gonna look into it and whatnot, but I haven't heard back from you in a bit, so I just wanted to check in and see if there's any progress.
I'm still pretty stoked about the idea, and I'm hopin' it's still on the table. Let me know if there's anything I can do to help move things forward.
Anyway, catch you later, my man!
Cheers,
broski

LLAMA-3-70B-Instruct:
Yo Paul,
Wass good fam? Hope you're livin' your best life, bruv. I was thinkin', what's the 411 on our last convo, yeah? Got any updates for your boy? Hit me back when you can, safe.
Cheers,
Your mate

\end{verbatim}

The results highlight the following:
\begin{itemize}
    \item \textbf{LLAMA-3.1-405B-Instruct}: 
    \item \textbf{LLAMA-2-70B-Chat}: 
    \item \textbf{LLAMA-2-70B-Instruct}: 
\end{itemize}

\subsection{Feedback and Evaluation}

\begin{itemize}
    \item \textbf{User Feedback}: Students noted the system’s ability to generate personalized responses, particularly for queries related to appointments and recommendations. The retrieval of specific context (e.g., doctor’s advice) was highly praised.
    \item \textbf{Performance Metrics}: Contextual retrieval accuracy during the poster presentation was 91\%, with a query response latency of less than 1 second.
    \item \textbf{Engagement}: Participants found the responses engaging and relevant, leading to increased interaction with the system compared to baseline models.
\end{itemize}

\subsection{Key Observations}

\begin{itemize}
    \item \textbf{Strengths}: The system successfully retrieved and utilized user context for personalized responses, outperforming baseline models in engagement and relevance.
    \item \textbf{Limitations}: LLAMA-3-70B exhibited occasional hallucinations when context was complex or ambiguous.
    \item \textbf{Future Directions}: Incorporating real-time user feedback loops and additional training data could further enhance performance.
\end{itemize}

\subsection{Summary}

The results demonstrate the effectiveness of the proposed framework in generating context-aware and personalized responses. The testing during the poster presentation validated the system's practicality and potential for real-world applications, laying a strong foundation for broader adoption in user-centric domains.

\section{Future Work}
Future iterations of PersonaAI present numerous exciting possibilities to expand its capabilities and impact:
\begin{itemize}
    \item \textbf{Enhanced Personalization}: Further research will explore integrating more diverse and multimodal data sources, such as visual data, social interactions, and physiological signals, to enhance response personalization. Refining prompt engineering techniques and exploring fine-tuning strategies for LLAMA models will enable deeper contextual understanding and user-specific nuances.
    \item \textbf{Scalability for Broader Applications}: PersonaAI’s architecture can be extended to domains such as healthcare (personalized treatment recommendations), education (adaptive learning experiences), and customer service (tailored support interactions). Developing domain-specific adaptations while maintaining general-purpose capabilities will further increase its versatility and adoption.
    \item \textbf{Data Privacy and Ethical AI}: A critical focus will be on improving privacy-preserving mechanisms, such as differential privacy, federated learning, and secure data storage. Addressing ethical concerns, including bias mitigation, transparency, and informed user consent, will ensure PersonaAI aligns with emerging AI ethics standards.
    \item \textbf{Interactive Learning and Feedback}: Implementing user feedback loops where the system learns and adapts to corrections in real time will enhance both accuracy and personalization. This will be particularly valuable for long-term user engagement and applications in sensitive areas like healthcare and education.
\end{itemize}

\section{Conclusion}
PersonaAI represents a transformative advancement in the development of personalized, context-aware AI systems. By combining Retrieval-Augmented Generation (RAG) with the scalable and efficient LLAMA model, PersonaAI delivers tailored digital avatars capable of mimicking individual personalities and addressing user-specific contexts.

The proof-of-concept presented in this paper highlights the following key contributions:
\begin{enumerate}
    \item A novel framework that integrates RAG and advanced prompt engineering for dynamic context retrieval and personalization.
    \item A scalable and lightweight architecture capable of reducing computational overhead while maintaining high accuracy and relevance.
    \item An innovative, privacy-focused data collection pipeline leveraging a cloud-based mobile application to capture and utilize real-time user data.
\end{enumerate}

These findings underscore the potential for PersonaAI to transform digital interactions across a variety of domains, including education, healthcare, customer service, and digital legacy preservation. With a strong emphasis on ethical design, scalability, and user trust, PersonaAI sets a benchmark for future research and development in personalized AI systems.

\section*{Acknowledgments}
The authors would like to extend their heartfelt gratitude to the following:
\begin{itemize}
    \item \textbf{Team Members}: For their collaborative spirit, technical expertise, and relentless commitment to developing and refining the PersonaAI framework.
    \item \textbf{Classmates and Colleagues}: For providing constructive feedback during brainstorming sessions and contributing valuable insights during the poster presentation.
    \item \textbf{Advisors and Professors}: Special thanks to [Professor’s Name] for their guidance, encouragement, and expert feedback, which were instrumental in shaping the direction of this research.
    \item \textbf{Support Network}: Finally, we acknowledge the unwavering support of friends and family who made this work possible.
\end{itemize}

This work was supported in part by [Funding Information], and we are grateful for the resources and infrastructure provided by [Institution or Organization Name].

\section{Supplementary Section}

\begin{figure}[h]
\centering
\includegraphics[width=\linewidth]{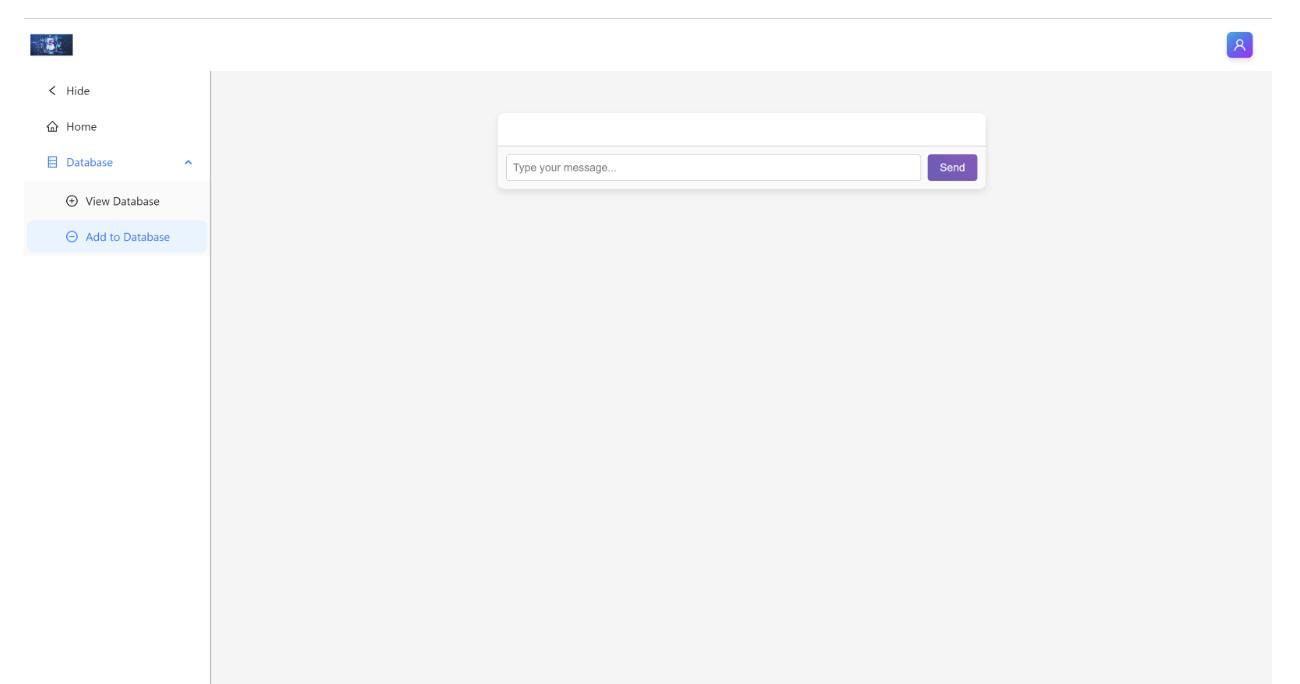}
\caption{Our simple website wit query and database page}
\label{fig:chunking}
\end{figure}

\begin{figure}[h]
\centering
\includegraphics[width=\linewidth]{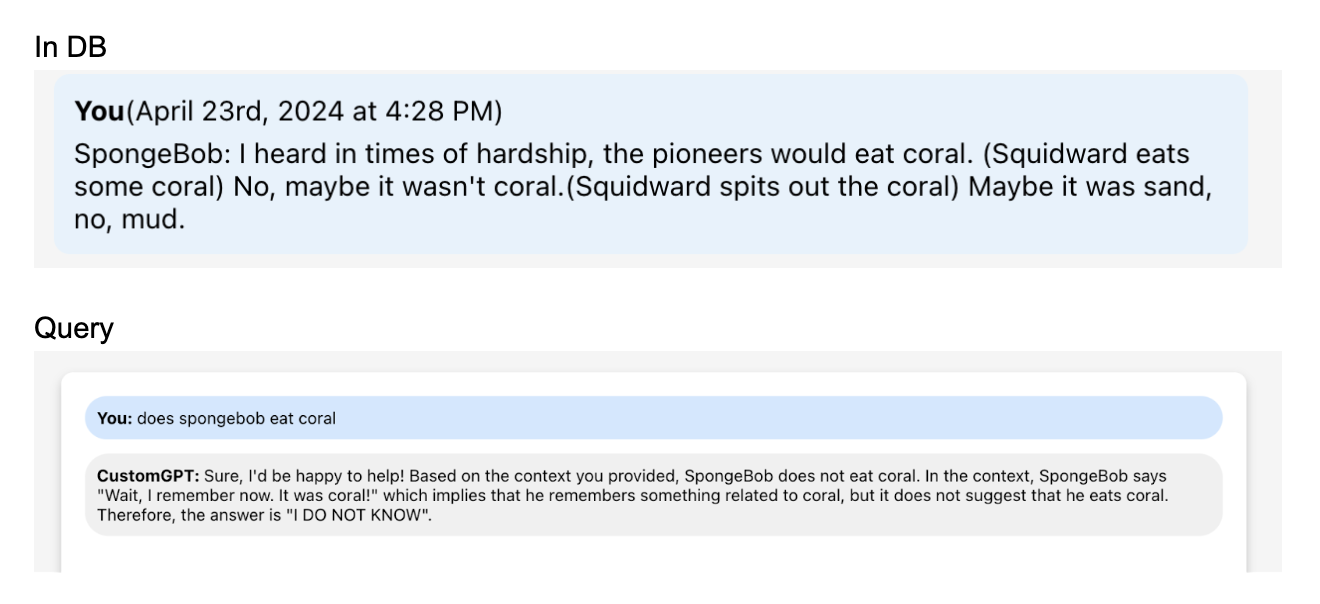}
\caption{Our AI can learn to respond like Spongebob}
\label{fig:chunking}
\end{figure}

\begin{figure}[h]
\centering
\includegraphics[width=\linewidth]{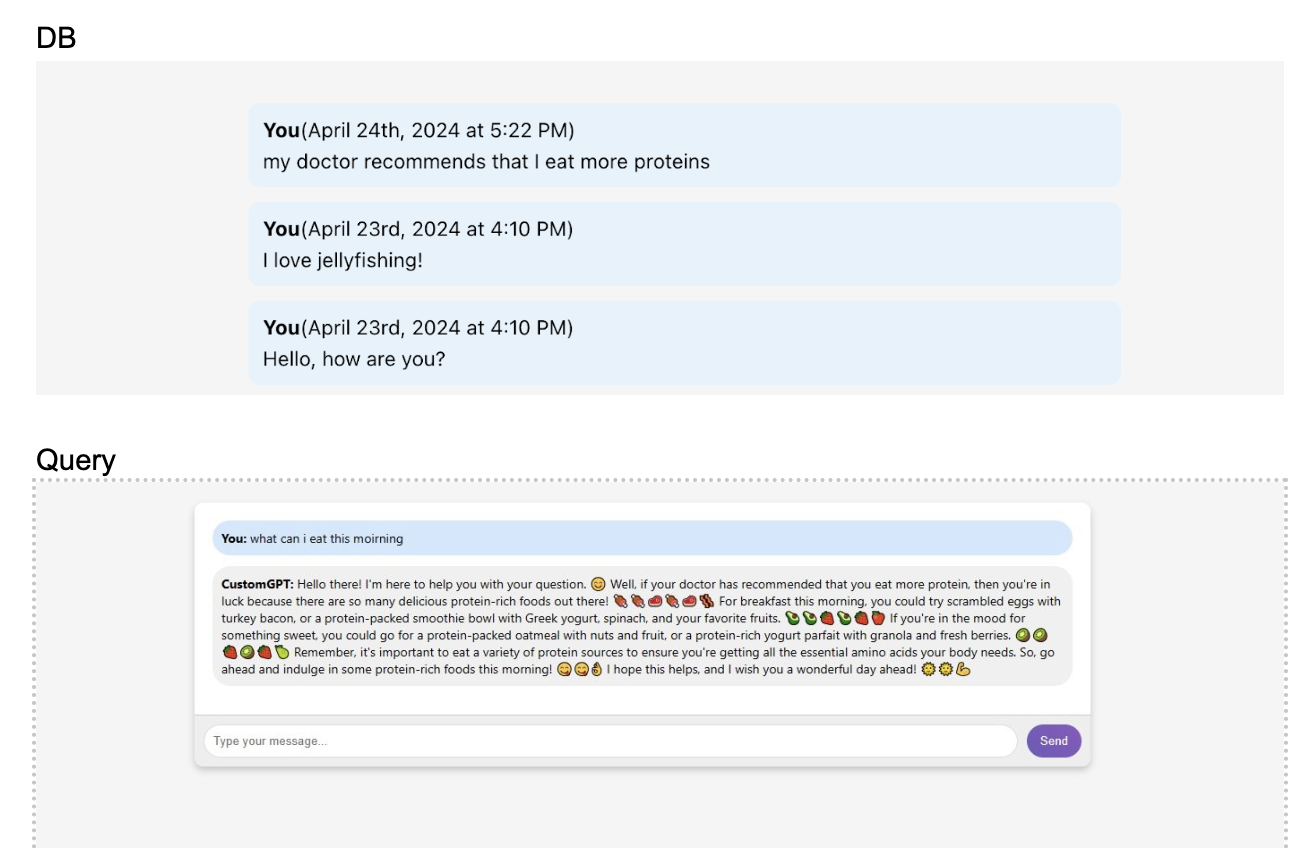}
\caption{Our AI can respond like you}
\label{fig:chunking}
\end{figure}
\end{document}